\documentclass[11pt, a4paper]{article}

\usepackage{jheppub}
\usepackage{amsfonts}
\usepackage{amsthm}
\usepackage{braket}
\usepackage{graphicx}
\usepackage{color}
\usepackage{dsfont}
\usepackage{verbatim}
\usepackage{amscd}
\usepackage{multirow}
\usepackage{subfigure}
\usepackage{tikz}
\usepackage{lipsum}  
\usepackage{kantlipsum}  
\usepackage{comment}

\usepackage{ bbold }

\usepackage{commath}
\usepackage{slashed}
\usepackage{braket}

\newcommand{\be}{\begin{equation}}
	\newcommand{\ee}{\end{equation}}
\newcommand{\bea}{\begin{eqnarray}}
	\newcommand{\eea}{\end{eqnarray}}

\newcommand{\eff}{\ensuremath{\mathrm{eff}}}

\begin{document}

\title{The chiral Gross-Neveu model on the lattice via a Landau-forbidden phase transition}

\begin{abstract}
{We study the phase diagram of the $(1+1)$-dimensional Gross-Neveu model with both $g_x^2(\bar{\psi}\psi)^2$ and $g_y^2(\bar{\psi}i\gamma_5\psi)^2$ interaction terms on a spatial lattice. The continuous chiral symmetry, which is present in the continuum model when $g_x^2=g_y^2$, has a mixed 't~Hooft anomaly with the charge conservation symmetry, which guarantees the existence of a massless mode. However, the same 't~Hooft anomaly also implies that the continuous chiral symmetry is broken explicitly in our lattice model. Nevertheless, from numerical matrix product state simulations we find that, for certain parameters of the lattice model, the continuous chiral symmetry reemerges in the infrared fixed point theory, even at strong coupling. We argue that, to understand this phenomenon, it is crucial to go beyond mean-field theory (or, equivalently, beyond the leading order term in a $1/N$ expansion). Interestingly, on the lattice, the chiral Gross-Neveu model appears at a Landau-forbidden second order phase transition separating two distinct and unrelated symmetry-breaking orders. We point out the crucial role of two different 't Hooft anomalies or Lieb-Schultz-Mattis obstructions for this Landau-forbidden phase transition to occur.
}
\end{abstract}

\author[a]{Gertian Roose,}
\author[a]{Jutho Haegeman,}
\author[a]{Karel Van Acoleyen,}
\author[a]{Laurens Vanderstraeten,}
\author[a,b]{and Nick Bultinck}

\affiliation[a]{Department of Physics and Astronomy, University of Ghent, Krijgslaan 281, 9000 Gent, Belgium}
\affiliation[b]{Rudolf Peierls Centre for Theoretical Physics, Parks Road, Oxford, OX1 3PU, UK}

\maketitle	
\flushbottom

\section{Introduction}

Discretizing quantum field theory (QFT) on a lattice in space or spacetime has been a very successful strategy to study interacting quantum fields using computational methods. The prevalent approach for the last decades has been to study the partition function of a quantum field theory, often including interacting gauge fields, on a spacetime lattice using some kind of Monte Carlo sampling. Indeed, the research field of lattice gauge theory has been tremendously successful in explaining the hadron masses and various other equilibrium properties of the standard model \cite{
creutz1983quarks, montvay1997quantum, rothe2012lattice}. More recently, there has been a renewed interest in quantum fields on a spatial lattice, either for classical simulation using the formalism of tensor networks, but also for quantum simulation using cold atoms or other discrete or analogue quantum simulators \cite{banuls2020simulating,meurice2020tensor}.

While the lattice (both in space and in spacetime) has the advantage of regularizing the divergences that typically occur as a result of the infinitely many degrees of freedom in a QFT, it is well known that certain symmetries of the field theory cannot be realized exactly in the lattice description. The most notorious example is that of chiral symmetry, which is a continuous U(1) symmetry of the massless Dirac operator in even spacetime dimensions. Even the discrete $\mathbb{Z}_2$ subgroup of the chiral symmetry cannot be implemented as an on-site symmetry in a local lattice model without causing a doubling of the number of Dirac fermions, a result known as (or resulting from) the Nielsen-Ninomiya theorem \cite{NielsenNinomiya}. By staggering the components of the Dirac spinor, it is possible to remove some of the doublers (and in particular all of them when only discretizing space in a (1+1)-dimensional theory) \cite{KogutSusskind}. The staggered model still breaks the full continuous chiral symmetry, but a single-site shift in the direction of staggering behaves as a discrete chiral transformation in the low-energy limit. The difficulty of realizing the chiral symmetry on the lattice is a consequence of the mixed 't Hooft anomaly \cite{tHooft} between the chiral and charge U(1) symmetries \cite{Wen}. Upon gauging the U(1) charge symmetry this 't Hooft anomaly gives rise to the Adler-Bell-Jackiw anomaly \cite{Adler,BellJackiw}, i.e.\ after gauging the current associated with the continuous chiral symmetry is no longer preserved.

In this work, we study the generalized Gross-Neveu (GGN) model in 1+1 dimensions \cite{GN}. The GGN model consists of $N$ massless Dirac fermions interacting via two different interaction terms $g_x^2(\bar{\psi}\psi)^2$ and $g_y^2(\bar{\psi}i\gamma_5\psi)^2$. When $g_x^2=g_y^2$, the interaction terms preserve the continuous chiral symmetry of the massless Dirac operator. Along this line with equal couplings, Coleman's theorem rules out the possibility that the chiral symmetry is broken spontaneously in $(1+1)$-dimensions. But even despite the absence of Goldstone modes, the mixed anomaly between the chiral symmetry and the charge conservation symmetry implies that the theory cannot be trivial in the infrared and must host a massless mode \cite{tHooft}. Everywhere away from the special line $g_x^2=g_y^2$ with continuous chiral symmetry, the remaining discrete chiral symmetry in the GGN Lagrangian is broken spontaneously, just as in the conventional GN model. The main question we address here is how much of these features of the continuum GGN model survive after discretizing the theory on a spatial lattice. Given that many properties of the continuum GGN phase diagram crucially hinge on the chiral symmetry and its 't Hooft anomaly, it is a priori not clear that a lattice discretized model --which breaks the chiral symmetry explicitly due to that same 't~Hooft anomaly-- will reproduce the continuum phase diagram (both at small and large coupling). 

Our analysis starts with a mean-field or large-$N$ calculation, which produces two different phase diagrams for the continuum and the lattice model, but suggests that fluctuations beyond mean-field theory (or subleading terms in the $1/N$ expansion) could be able to remove the apparent discrepancy. A fully unbiased matrix product state simulation for the $N=2$ lattice GGN model confirms this expectation, and produces a phase diagram which contains a critical line that has the same infrared behaviour as the chiral GN model. This critical line appears as a Landau-forbidden second order phase transition of the lattice model which separates two gapped phases with unrelated spontaneously broken discrete symmetries. We argue that this Landau-forbidden phase transition can occur as a critical \emph{line} in the lattice model due to the presence of \emph{two} different Lieb-Schultz-Mattis (LSM) obstructions \cite{LSM,HastingsLSM,OshikawaLSM}, which are lattice versions of the continuum 't Hooft anomalies. One of these LSM obstructions is related to a lattice version of the mixed 't Hooft anomaly between the remaining discrete chiral symmetry and the charge conservation symmetry. The other LSM obstruction is less well-known, and it relies on a combination of several different symmetries including charge conjugation and spatial reflection symmetry.

The paper is structured as follows. In the following section, we start by providing a short review of the (chiral) GN model. More specifically, we highlight some often overlooked symmetries of the model and use bosonization to provide a nonperturbative argument for the existence of a critical line in the phase diagram. In the same section we introduce the lattice model based on the symmetries that are present in the continuum. Section III presents the mean field solution, which coincides with the large-$N$ limit, for both the continuum and the lattice model, and discusses its shortcomings. In Section IV we use tensor network methods to determine the phase diagram of the $N=2$ lattice model. The phase diagram exhibits a critical line between two symmetry broken phases, which we can identify with the chiral GN QFT in the continuum limit. In Section V, we reinterpret our lattice model from a condensed matter perspective to further discuss the nature of our critical line in the context of the LSM theorem. Section VI summarises our main conclusions.

\section{The generalized Gross-Neveu model}

We study the generalized Gross-Neveu model \cite{GN} with $N$ flavors, which in the continuum is described by the following action:
\begin{equation}\label{action}
    S = \int \mathrm{d}x\mathrm{d}t\left( \sum_c \bar{\psi}_ci\slashed{\partial}\psi_c + \frac{g_x^2}{2N}\del{\sum_c\bar{\psi}_c\psi_c}^2 + \frac{g_y^2}{2N}\del{\sum_c\bar{\psi}_ci\gamma_5\psi_c}^2 \right)\,,
\end{equation}
where $\psi_c$ is the two component Dirac spinor for each of the flavors $c=1,\ldots,N$. The matrices $\gamma^\mu$ satisfy the usual Clifford algebra $\{\gamma^\mu,\gamma^\nu\}=2\eta^{\mu\nu}$ (we use $\eta = \text{diag}(1,-1)$) and are used to define $\bar{\psi}_c = \psi^\dagger_c \gamma^0 $, $\slashed \partial = \gamma^\mu \partial_\mu$, and $\gamma_5 = \gamma^0\gamma^1$. In the remainder of this section we review all the symmetries of this action, discuss the phase diagram and re-express the action in terms of bosonic fields.

\subsection{Review of the continuum symmetries}\label{reviewGN}

In the general case $g_x^2 \neq g_y^2 \neq 0$, the relevant internal symmetries are:\footnote{The various symmetries also interact. Charge conjugation flips the rotation angle of charge U(1) (and combines with it into an $O(2)$ group) as well as of chiral (axial) rotation. The discrete chiral transformation also anticommutes with the spatial reflection (toghether they generate the Pauli group).}
\begin{align}
    &\text{SU($N$)} \text{ flavor rotation : }& \psi_c&\rightarrow U_{cc'}\psi_{c'} \\
    &\text{U(1) charge rotation : }& \psi_c &\rightarrow e^{i\theta}\psi_c \nonumber \\ 
    &\mathbb{Z}_2^{\mathcal{D}}\text{ discrete chiral transformation : }& \psi_c&\rightarrow \gamma_5 \psi_c  \nonumber  \\
    &\mathbb{Z}_2^{\mathcal{C}}\text{ charge conjugation : }&\psi_c &\rightarrow \gamma_C\psi_c^\ast, \nonumber 
\end{align}
where the unitary matrix $\gamma_C$ is defined such that $\gamma_C^\dagger\gamma^0 \gamma_C = -(\gamma^0)^{\mathsf{T}}$ and $\gamma_C^\dagger\gamma_5 \gamma_C = (\gamma_5)^{\mathsf{T}}$. Besides these internal symmetries, the action naturally has spacetime symmetries, namely the full Poincar\'{e} group, which includes Lorentz transformations, spacetime translations, spatial reflection and time reversal. From these, we only highlight the reflection symmetry, which acts as
\begin{equation}
    \mathbb{Z}_2^{\mathcal{R}} \text{ spatial reflection} : \psi \rightarrow \gamma_R\psi\,,\;\;\; x\rightarrow -x\, ,
\end{equation}
where $\gamma_R$ satisfies $\gamma_R^\dagger \gamma^0 \gamma_R = \gamma^0$ and $\gamma_R^\dagger \gamma_5 \gamma_R = -\gamma_5$. A typical choice is $\gamma_R = \gamma^0$.

Let us now make a particular basis choice, such that the gamma matrices are given by the Pauli matrices $\gamma_5 = \sigma^x$ and $\gamma^0 = \sigma^y$. In this basis we find that $\gamma_C =  \mathbb{1}$ and $\gamma_R = \sigma_y$. The bilinears $\bar{\psi}\psi$ and $\bar{\psi} i\gamma_5\psi$ transform respectively as scalar and pseudoscalar quantities with respect to both reflection and charge conjugation, whereas both or of course pseudoscalars with respect to the discrete chiral tranformation.

For $g_y^2 = 0$, the charge conjugation action $\mathcal{C}$ can be extended to a $\mathbb{Z}_2^{\otimes N}$ symmetry by applying it to each flavor separately. Consequently, the $\mathbb{Z}_2^{\otimes N}$, U(1) and SU($N$) symmetries can be embedded in a larger O(2$N$) symmetry, which can be made manifest by rewriting the complex Dirac fermions $\psi_c$ in terms of two real Majorana fermions: $\psi_c = (\chi_{2c-1}+i\chi_{2c})/\sqrt{2}$, where $\chi_c^\dagger = \chi_c$ and $\{\chi_c,\chi_{c'}\} = 2\delta_{c,c'}$. Similarly, an enhanced O(2$N$) symmetry is also present when $g_x^2=0$. This O(2$N$) group now contains the $\mathcal{DC}$ symmetry action (generating a $\mathbb{Z}_2^{\otimes N}$ symmetry group when $g_x^2=0$), and again the U(1) and SU($N$) symmetry groups. The O(2$N$) symmetry at $g_x^2=0$ becomes explicit after rewriting the complex Dirac fermions $\psi'_c = \exp(i\pi\sigma^x/4)\psi_c$ in terms of two real Majorana fermions: $\psi_c' = (\chi'_{2c-1}+i\chi'_{2c})/\sqrt{2}$.

Finally, when the two interaction coefficients $g_x^2$ and $g_y^2$ are equal, the generalized Gross-Neveu model is known as the `chiral Gross-Neveu model', which can be interpreted as a (1+1)-dimensional version of the `Nambu-Jona-Lasinio' model \cite{NJL1,NJL2}. Here the chiral symmetry becomes continuous, i.e.\ 
\begin{equation}\label{U1A}
    \text{U}_A(1) : \psi_c \rightarrow e^{i\theta_A\gamma_5}\psi_c\,
\end{equation}
becomes a symmetry of the action.

\subsection{Phase diagram and bosonization}

The phase diagram of the generalized Gross-Neveu (GN) model is well-understood, and we review it here. First, for $g_y^2=0$, the action reduces to that of the conventional GN model. In this case, the interaction leads to dynamical mass generation for the Dirac fermions, spontaneously breaking the discrete chiral symmetry, which is characterized by the fact that the vacuum obtains a chiral condensate: $\langle \bar{\psi}\psi\rangle \neq 0$. For $g_x^2=0$, the situation is analogous to that of the conventional GN model, as we can transform the $g_x^2$ and $g_y^2$ interaction terms into each other with a chiral symmetry rotation, where now the chiral condensate is characterized as $\langle \bar{\psi}i\gamma_5\psi\rangle \neq 0$. As long as $g_x^2\neq g_y^2$, the IR physics does not change if we move away from the lines with either $g_x^2=0$ or $g_y^2=0$. In particular, for $g_x^2>g_y^2$, the dynamical mass generation is associated with a chiral condensate $\langle \bar{\psi}\psi\rangle \neq 0$, whereas $\langle \bar{\psi}i\gamma_5\psi\rangle$ remains zero (and vice versa for $g_x^2<g_y^2$).

Along the $g_x^2 = g_y^2$ chiral line, the IR physics drastically changes due to the presence of the continuous chiral symmetry, which is a proper symmetry at the quantum level, as no gauge fields are included. The Coleman-Hohenberg-Mermin-Wagner (CHMW) theorem \cite{Coleman,Hohenberg,MerminWagner} excludes spontaneous breaking of this continuous chiral symmetry, which automatically implies that both $\langle \bar{\psi}\psi\rangle = 0$ and $\langle \bar{\psi}i\gamma_5\psi\rangle = 0$. However, despite the fact that there is no chiral condensate along the line $g_x^2=g_y^2$, the Dirac fermions nevertheless acquire a dynamically generated mass.

To better understand the mechanism responsible for dynamical mass generation along (and close to) the line $g_x^2 = g_y^2$, it is insightful to consider the bosonized version of Eq.~\eqref{action}. Here, we only consider the $N = 2$ case, both for simplicity of the presentation and because this is also the model that we study numerically (for details of the bosonization procedure for general $N$, we refer to \cite{Ha_2013}). Bosonization allows us to map the fermion action to a theory of two compact bosons $\phi_1$ and $\phi_2$ with compactification radius $2\pi$. Under this mapping, the kinetic term becomes 
\begin{align}\label{bosonkin}
    \bar{\psi}_1 i\slashed\partial\psi_1 + \bar{\psi}_2 i\slashed\partial\psi_2 &\rightarrow \frac{1}{8\pi} \left[ (\partial_\mu \phi_1)^2 + (\partial_\mu \phi_2)^2\right]\,,
\end{align}
and the chiral transformation $\psi_c \rightarrow e^{i\theta_A\gamma_5}\psi_c$ corresponds to a shift of the scalar fields: $\phi_c \rightarrow \phi_c + \theta_A$. The mappings for fermion bilinears are:
\begin{align}
\begin{cases}
    \bar{\psi}_1\psi_1 + \bar{\psi}_2\psi_2 &\rightarrow -\frac{1}{\alpha}(\cos\phi_1 + \cos \phi_2) \\
    \bar{\psi}_1i\gamma_5\psi_1 + \bar{\psi}_2i\gamma_5\psi_2 &\rightarrow \phantom{-}\frac{1}{\alpha}(\sin\phi_1 + \sin\phi_2) \,,
\end{cases}
\label{bos dic}
\end{align}
where $\frac{1}{\alpha}$ is a UV-cutoff. Using these relations, we arrive at the following bosonized action:
\begin{align}
    S &= \int \mathrm{d}^2x \frac{1}{8\pi }\left[(\partial_\mu \phi_1)^2 + (\partial_\mu \phi_2)^2\right] + \frac{g_x^2}{4\alpha^2} \del{\cos{\phi_1}+\cos{\phi_2} }^2 + \frac{g_y^2}{4\alpha^2}\del{\sin{\phi_1}+\sin{\phi_2}}^2 \\
      &= \int \mathrm{d}^2x \frac{1}{8\pi }\left[(\partial_\mu \phi_1)^2 + (\partial_\mu \phi_2)^2\right]  + \frac{g^2_x+g_y^2}{4\alpha^2}\cos(\phi_1 - \phi_2) \nonumber \\
      & \hspace{5cm}  + \frac{g_x^2-g_y^2}{4\alpha^2}\cos(\phi_1+\phi_2)\left(1+\cos(\phi_1-\phi_2)\right)\,. \nonumber
\end{align}
If we now write the boson fields as $\phi_1 = \theta + \varphi$ and $\phi_2 = \theta - \varphi$, then the bosonized action takes on a particularly simple form:
\begin{multline}
    S = \int \mathrm{d}^2x \frac{1}{2\pi K}(\partial_\mu \theta)^2+ \frac{1}{2\pi K}(\partial_\mu \varphi)^2  +\frac{g_x^2+g_y^2}{4\alpha^2}\cos(2\varphi) \\ + \frac{g_x^2-g_y^2}{4\alpha^2}\cos(2\theta)(1+\cos(2\varphi))\, ,\label{bosonic_critical}
\end{multline}
where $K = 2$. Along the line with continuous chiral symmetry, i.e. when $g_x^2 = g_y^2$, this action describes one interacting boson $\varphi$, which transforms trivially under the chiral U(1) symmetry, and one free boson $\theta$, which transforms as $\theta\rightarrow \theta + \theta_A$. Although the $\cos(2\varphi)$ is marginal at the classical level for $K = 2$, it becomes relevant at the quantum level by renormalizing $K$ to smaller values (this can be seen from the Kosterlitz RG equations \cite{Kosterlitz}). As a result, the $\cos(2\varphi)$ term causes the $\varphi$ field to condense. 

In the chiral GN model the compact boson $\theta$ is gapless because not only does the chiral U(1) symmetry forbid terms of the form $\cos(n\theta)$, $\theta$ can also not be disordered by proliferating vortices (i.e. instantons which change the winding of $\theta$). The reason is that the charge current in the presence of a spatially varying $\theta$ configuration, relative to the charge current of the vacuum, is given by the Goldstone-Wilczek formula \cite{GoldstoneWilczek}:
\begin{equation}
    J_\mu = \frac{2}{2\pi}\epsilon_{\mu\nu}\partial_\nu \theta \,.
\end{equation}
From this relation we see that the electric charge corresponds to the winding of $\theta$ along the spatial direction: $Q = \int\mathrm{d}x\, \partial_x\theta/\pi$. As a consequence, vortices in $\theta$ are forbidden by the charge conservation, i.e.\ by the U(1) charge symmetry. This is a manifestation of the `t Hooft anomaly, which rules out a trivial IR fixed point if both the charge and chiral U(1) symmetries are to be preserved. We thus arrive at the conclusion that the IR fixed point of the chiral GN model is a single compact boson. This conformal field theory has a central charge $c = 1$, instead of $c = 2$ as for two free Dirac fermions ($g^2_x = g^2_y = 0$). This is a manifestation of the fact that the fermions have acquired a mass.

When moving away from the line with equal couplings the continuous chiral symmetry breaks down to the discrete chiral symmetry $\mathbb{Z}_2^\mathcal{D}$. From Eq.~\eqref{bosonic_critical}, we see that the effective action describing the IR physics close to the chiral line is
\begin{align}
    S = \int \mathrm{d}^2x \frac{1}{2\pi K} (\partial_\mu \theta)^2 + \delta \cos 2\theta\, ,
    \label{IR_action}
\end{align}
where we have introduced $\delta = (g_x^2-g_y^2)/4\alpha^2$ and we have dropped an irrelevant term. The $\cos(2\theta)$ term in Eq.~\eqref{IR_action} is relevant for the same reason that the $\cos(2\varphi)$ discussed above is relevant. In Sec. \ref{MPS simul} we will show that we can recover the IR physics described by \eqref{IR_action} by simulating the GGN on the lattice, even though we cannot preserve the chiral symmetry explicitly. As we will see below, one consequence of the loss of continuous chiral symmetry is that the relation $\delta = (g_x^2-g_y^2)/4\alpha^2$ no longer holds for the parameters of our lattice model, which we now introduce.

\subsection{Lattice model}
Let us now introduce the specific lattice discretization of the GGN that we will study. We use a particular realization of the standard staggered fermion discretization \cite{KogutSusskind,Roose2020}, where the two components of the Dirac fermions are defined to live on neighbouring lattice sites. The free/kinetic part of the Hamiltonian is obtained by using a symmetric finite difference approximation for the spatial derivative, and by using the same basis choice ($\gamma_5 = \sigma^x$ and $\gamma^0 = \sigma^y$) as in the previous section. In this way, we arrive at the following kinetic or hopping term on the lattice:

\begin{eqnarray}
H_K & = & a^{-1}\sum_n K_{n,n+1}\\
    & = & -ia^{-1}\sum_c\del{\varphi_{c,n}^\dagger \varphi_{c,n+1} - \varphi_{c,n+1}^\dagger \varphi_{c,n}} \, ,
\end{eqnarray}
where $n$ ($c$) labels the lattice sites (flavors), $a$ is the lattice constant, and $\varphi^\dagger_{c,n}$ and $\varphi_{c,n}$ are fermionic creation and annihilation operators satisfying $\{\varphi^\dagger_{c,n},\varphi^\dagger_{c',n'}\} = \{\varphi_{c,n},\varphi_{c',n'}\}=0 $ and $\{\varphi_{c,n},\varphi^\dagger_{c',n'} \} = \delta_{c,c'}\delta_{n,n'}$. The kinetic term admits two different mass terms, which, with our basis choice, are given by 

\begin{eqnarray}
m\bar\psi \psi \rightarrow m\psi^\dagger \sigma^y \psi & \rightarrow & ma (-1)^n \left(\frac{K_{n-1,n}-K_{n,n+1}}{2}\right) \label{mass1}\\
m\bar\psi i\gamma_5 \psi \rightarrow m\psi^\dagger \sigma^z \psi & \rightarrow & ma (-1)^n \left(O_{n}-O_{n+1}\right) \label{mass2}\,, 
\end{eqnarray}
with $O_n = \sum_c \varphi^\dagger_{c, n} \varphi_{c,n}$. Note that both mass terms are odd under a translation by one lattice site, as expected from the fact that a single-site translation should behave as the discrete chiral transformation in the low-energy limit. The first mass term $m\psi^\dagger \sigma^y\psi$ translates on the lattice to a bond order parameter, which promotes dimerization on even or odd lattice bonds, whereas the second mass term $m\psi^\dagger \sigma^z\psi$ results in a polarization of the lattice fermions on either the even or odd lattice sites, i.e.\ it creates an imbalance between the average occupation of the even and odd lattice sites.

For the discretized interaction terms, we simply take the squares of both possible mass terms/order parameters. The final lattice Hamiltonian then takes on the following form:
\begin{align}
	H =  a^{-1}\sum_n \del{  K_{n,n+1} - \frac{g_x^2}{4N}\del{\frac{K_{n,n+1}-K_{n+1,n+2}}{2}}^2 - \frac{g_y^2}{4N}\del{O_n-O_{n+1}}^2  } \,.
	\label{latticeH}
\end{align}
This Hamiltonian manifestly preserves the internal U(1), SU($N$), $\mathbb{Z}_2^{\mathcal{C}}$ and $\mathbb{Z}_2^{\mathcal{D}}$ symmetries of the continuum model, as well as the spatial translation and reflection symmetries. As mentioned above, the discrete chiral symmetry of the QFT does not act as an exact internal symmetry, but can be related to one-site spatial translations $\mathcal{T}$ in the low-energy limit. Regarding the reflection symmetry, it should be noted that the lattice exhibits two possible reflection transformations, namely across bonds and across sites. From the form of $\gamma_R$ in the reflection in the continuum, it can be noted that it interchanges the two components of the Dirac spinor. As we are using the staggered formulation, this should amount to interchanging even and odd sites on the lattice, which corresponds to a bond-centered reflection. A bond-centered reflection $n \to 1-n$ in itself maps $K_{n,n+1}$ to $-K_{-n,-n+1}$, so we also need to add a local action, such that the $\phi_n$ operators on neighbouring sites acquire an opposite sign. A local charge rotation $\exp(i n \pi \sum_{c} \phi_{c,n}^\dagger \phi_c)$ (which acts as the identity every second site) accomplishes this goal. Below, we denote with $\mathcal{R}_B$ this bond centered reflection, including the additional on-site action. A site-centered reflection (including the same on-site action) can be interpreted as $\mathcal{T}\mathcal{R}_B$, or thus as the combination of a discrete chiral transformation and a reflection.

For our MPS simulations, we further transform the lattice fermion Hamiltonian in Eq.~\eqref{latticeH} into a lattice spin Hamiltonian via a Jordan-Wigner transformation, where each fermion operator is represented in terms of Pauli matrices as
\begin{equation}
    \varphi_{c,n} = \left(\prod_{n'<n} \prod_{c'}\sigma^z_{c',n'} \right)\left( \prod_{c'<c} \sigma^z_{c',n} \right)\sigma^-_{c,n}\, ,
\end{equation}
where $\sigma^- = (\sigma^x - i\sigma^y)/2$, and we have introduced a linear ordering for the different flavors. In a previous work \cite{Roose2020}, we have numerically studied the two-flavor version of the lattice Hamiltonian in Eq.~\eqref{latticeH} with $g_y^2 = 0$ using MPS. We were able to take the continuum limit of our numerical results and recover some of the QFT results to very high accuracy, thus confirming the validity of both our lattice Hamiltonian and our MPS methods.

Before concluding our discussion of the lattice Hamiltonian, let us point out a subtlety about the O(2$N$) symmetries which are present in the continuum action when either $g_y^2=0$ or $g_x^2=0$. If $g_y^2=0$, then the full O(2$N$) symmetry of the continuum model is present in the lattice Hamiltonian, and acts in a local way. This is possible because the O(2$N$) symmetry group contains the $\mathcal{C}$ symmetry action, which acts locally in the lattice model, and generates a $\mathbb{Z}_2^{\otimes N}$ symmetry by acting on each flavor separately if $g_y^2=0$. Indeed, this was the motivation for our basis choice of the gamma matrices, where $\gamma_C = \mathbb{1}$. As explained in the previous section, the continuum model at $g_x^2=0$ also possesses an O(2$N$) symmetry, where now the $\mathbb{Z}_2^{\otimes N}$ subgroup is generated by acting with the $\mathcal{DC}$ on each flavor separately. The $\mathcal{DC}$ symmetry on the lattice, however, does not act locally as it contains a discrete chiral symmetry action $\mathcal{D}$, which we discussed above. As a result, there is no lattice analogue of acting with $\mathcal{DC}$ on a single fermion flavor. This implies that the duality for interchanging $g_x \leftrightarrow g_y$, which exist in the continuum and is generated by applying a $\pi/2$ chiral rotation, does not exist in the lattice model. Despite this shortcoming of the discretization, we argue below that our numerical results for $N = 2$ with both $g_x^2$ and $g_y^2$ non-zero agree well with the results expected from the continuum model.

\section{Large-$N$ solution}

In this section we analyse the GGN model in the large-$N$ limit, where mean-field theory becomes exact. In order to keep this paper self-contained, we first review the large-$N$ solution of the continuum model. We compare the solutions of the continuum and lattice theories, and discuss the implications of the broken continuous chiral symmetry on the lattice.
%%%%%%%%%%%%%%%%%%%%%%%%%%%%%%%
\subsection{Continuum model}

\label{largeN}
The Hamiltonian of the generalized Gross-Neveu model in the continuum is :
\begin{align}
    H = \int \mathrm{d}x \del{\bar{\psi}i\gamma^x \partial_x \psi - \frac{g_x^2}{2N}(\bar{\psi}\psi)^2 - \frac{g_y^2}{2N}(\bar{\psi}i\gamma_5\psi)^2   } \,,
\end{align}
where, as before, $\psi$ is a $2N$-component Dirac spinor. In taking the $N\rightarrow\infty$ limit we can exploit the monogamy of entanglement to write the ground state as a product state over the different flavors: $\ket{\Psi}=\ket{\phi}^{\otimes N}$ (see e.g. Ref. \cite{Finetti}). The energy per flavor of such states is given by :
\begin{align}
    \frac{E}{N} = \int \mathrm{d}x &\Braket{\bar{\psi}_si\gamma^x\partial_x \psi_s - \frac{g_x^2}{2N} (\bar\psi_s\psi_s )^2 - \frac{g_y^2}{2N} (\bar\psi_s i\gamma_5\psi_s )^2    } \nonumber \\
    &\hspace{2cm}- \frac{g_x^2}{2}\frac{N-1}{N} \Braket{\bar{\psi}_s\psi_s}^2 - \frac{g_y^2}{2}\frac{N-1}{N} \Braket{\bar{\psi_s}i\gamma_5\psi_s}^2  \,,
\end{align}
where $\psi_s$ is a $2$-component single-flavor Dirac spinor. For sufficiently large $N$ the terms proportional to the expectation values of the fluctuations, i.e. $\braket{ (\bar{\psi}_s \psi_s)^2 }$ and $\braket{(\bar{\psi}_s i\gamma_5 \psi_s)^2}$, can be neglected. Varying the energy with respect to the single-flavor wave function while using a Lagrange multiplier to ensure normalisation, gives the following eigenvalue problem:
\begin{align}
    \int \mathrm{d}x \del{\bar{\psi}_si\gamma^x\partial_x\psi_s - g_x^2 \sigma\hspace{1mm} \bar{\psi}_s\psi_s - g_y^2 \pi\hspace{1mm} \bar{\psi}_si\gamma_5 \psi_s } \ket{\phi} = H_{\text{MF}}\ket{\phi} = E_{\text{MF}} \ket{\phi} \,,
\end{align}
where $\sigma$ and $\pi$ respectively are the (translationally invariant) expectation values $\braket{\bar\psi_s \psi_s} $ and $\braket{\bar\psi_s i \gamma_5 \psi_s}$, such that these equations have to be solved self-consistently. For now we can easily diagonalize the effective mean-field Hamiltonian in momentum space and we find the following single-particle dispersion relation:
\begin{align}\label{cont_disp}
    \varepsilon_{\text{MF}}(k) = \pm \sqrt{k^2 + g_x^4 \sigma^2 + g_y^4 \pi^2} \,.
\end{align}
The groundstate $\ket{\Omega}$ of $H_{MF}$ simply corresponds to the filled Dirac sea of the states with negative energy. We define the effective potential as the energy density of $\ket{\Omega}$:
\begin{align}
    V_{\text{eff}}(\sigma, \pi) = \frac{g_x^2}{2}\sigma^2 + \frac{g_y^2}{2}\pi^2 - \int \frac{\mathrm{d}k}{2\pi} \sqrt{k^2 + g_x^4 \sigma^2 + g_y^4 \pi^2} \, .
    \label{effpot_cont}
\end{align}
Let us now introduce polar coordinates for the order parameters:
\begin{align}
\begin{cases}
\sigma = \rho \cos \theta \\
\pi = \rho \sin\theta\, ,
\end{cases}
\label{MF bos}
\end{align}
where we have used, not coincidentally, the same notation as in the bosonization formula \eqref{bos dic}. Indeed, under chiral transformations the $\theta$ field from Eq.~\eqref{MF bos} transforms identically to the $\theta$ field introduced in Eq.~\eqref{bosonic_critical}. Using the $\rho$ and $\theta$ variables, the effective potential can be written as
\begin{equation}
    V_{\text{eff}}(\rho,\theta) = \frac{g^2}{2}\rho^2 + \frac{\Delta g}{2}\rho^2\cos2\theta - \int \frac{\mathrm{d}k}{2\pi}\sqrt{k^2 + (g^4 + \Delta g^2)\rho^2 + 2g^2\Delta g\rho^2\cos2\theta}\, ,
\end{equation}
where $g^2 = (g_x^2+g_y^2)/2$ and $\Delta g = (g_x^2-g_y^2)/2$. Minimizing this effective potential (after introducing a cutoff $\Lambda$) is equivalent to solving the mean-field self-consistency equations. If $\Delta g \neq 0$, and assuming $\rho^2\neq 0$, one finds that the minima of $V_{\text{eff}}$ are located at either $\theta = 0,\pi$ or $\theta = \pm \pi/2$ because $V_{\text{eff}}$ depends only on $\theta$ via $\cos 2\theta$. Using this fact, we find from minimizing the effective potential with respect to $\rho^2$ that
\begin{equation}
    \rho^2 =
    \begin{cases}
    \frac{4\Lambda^2}{g_x^2}e^{-2\pi/g_x^2} & \text{ if } g_x^2\geq g_y^2 \\
    \frac{4\Lambda^2}{g_y^2}e^{-2\pi/g_y^2} & \text{ if } g_x^2\leq g_y^2\, ,
    \end{cases}
\end{equation}
such that $\sigma$ and $\pi$ are never simultaneously equal to zero, except when $g_x^2=g_y^2=0$. We are therefore led to the conclusion that the Dirac fermions acquire a mass for all non-zero values of the couplings. 

A non-zero value for $\rho$ also implies that the chiral symmetry is spontaneously broken. For the chiral GN model, however, this is an artefact of the large-$N$ limit, as the CHMW theorem implies that in $1+1$ spacetime dimensions fluctuations around mean-field theory will restore the continuous chiral symmetry at any finite $N$. However, although the chiral symmetry is restored beyond mean-field theory, the Dirac fermions nevertheless remain everywhere gapped. The physical picture is that, at finite $N$, the field $\rho^2 = \sigma^2+\pi^2$ retains a non-zero expectation value, thus providing a mass scale for the fermions, while at the same time, the long-range order for the $\theta$ field in mean-field theory is replaced with quasi-long range or algebraic order at finite $N$. The effective IR action describing these fluctuations is exactly the compact boson introduced previously in Eq.~\eqref{IR_action}.
%%%%%%%%%%%%%%%%%%%%%%%%%%%%%%%%%%%
\subsection{Lattice model}
%%%%%%%%%%%%%%%%%%%%%%%%%%%%%%%%%%5
\begin{figure}
	\centering
	\begin{minipage}{0.93 \textwidth}
           \vspace{0.5cm}
	       \includegraphics[width = \textwidth]{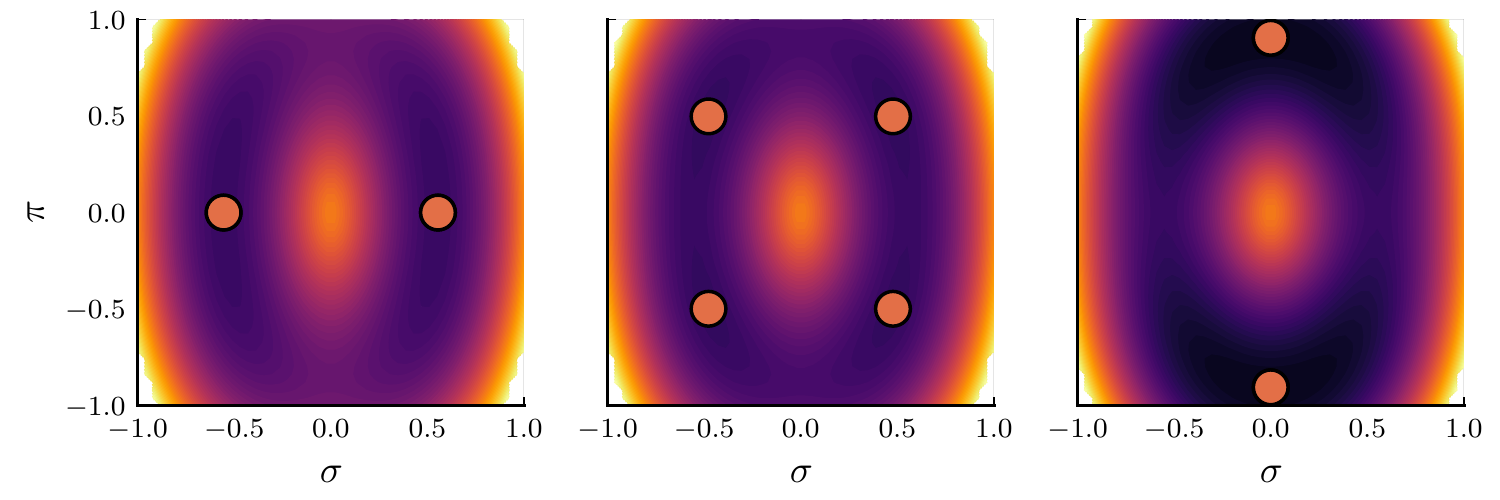}
	\end{minipage}
	\begin{minipage}{0.06 \textwidth}
	\includegraphics[width = \textwidth, trim = {5.9cm, 0, 0.5cm, 0}, clip]{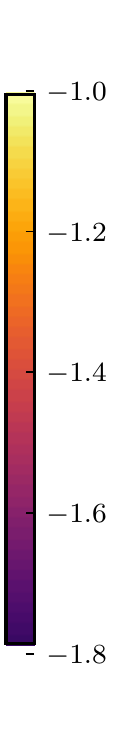}
	\end{minipage}
	\caption{\emph{(the colorscale and y-axis is the same for all three figures)} \\The effective potential $V_{\text{eff}}(\sigma,\pi)$ of the lattice model as a function of $\sigma$ and $\pi$ for three different combinations of the coupling constants.  All figures have $g_x^2=6.0$ but $g_y^2$ is taken from $2.6$ to $2.8$ and finally to $3.2$. These couplings are chosen so that the leftmost figure sits in the bond density wave phase, where the effective potential has two minima with nonzero $\sigma$. The second figure represents the coexistence region where both order parameters are nonzero. Note that in this coexistence region the effective potential is close to spherically symmetric even this far away from the continuum limit. Finally, we show the effective potential for a value of $g^2_y$ where the minima are found for nonzero $\pi$ and the groundstate has a sublattice-polarized fermion occupation.} \label{effective_potential}
\end{figure}

Let us next perform the mean-field analysis of our proposed lattice version of GGN model. Once again we exploit the monogamy of entanglement and calculate the energy of the groundstate with respect to the states $\ket{\Psi}=\ket{\phi}^{\otimes N}$ :
\begin{align}
    	\frac{\Braket{H}}{N} = \sum_n \Braket{ k_{n,n+1} }- \frac{g^2_x}{2}\Braket{\sigma_{n,n+1,n+2}}^2 - \frac{g^2_y}{2}\Braket{\pi_{n,n+1}}^2 \,,
\end{align}
where we have already neglected terms proportional to expectation values of fluctuations and introduced the following shorthand notations :
\begin{eqnarray}
    k_{n,n+1} & = & -i(\varphi^\dagger_{c,n}\varphi_{c,n+1} - \varphi^\dagger_{c,n+1}\varphi_{c,n}) \\
    \sigma_{n,n+1,n+2} & = & \frac{1}{2}(k_{n,n+1} - k_{n+1,n+2}) \\ 
    \pi_{n,n+1} & = & \varphi^\dagger_{c,n}\varphi_{c,n} - \varphi^\dagger_{c,n+1}\varphi_{c,n+1} \,.
\end{eqnarray}
Variation with respect to the single-flavor wave function while using a Lagrange multiplier to ensure normalisation, gives the following eigenvalue problem
\begin{align}
	\sum_n \del{k_{n,n+1} - g_x^2\braket{\sigma_{n,n+1,n+2}}\sigma_{n,n+1,n+2}  - g_y^2 \braket{\pi_{n,n+1}}\pi_{n,n+1}  }\ket{\phi} = E_{\text{MF}} \ket{\phi}\, ,
	\label{SC_hamiltonian}
\end{align}
similar to what we found in the continuum model. We are interested in states with a two-site unit cell. Consequently, we diagonalize \eqref{SC_hamiltonian} under the conditions that $\langle\sigma_{n,n+1,n+2}\rangle = (-1)^n \sigma$ and $\langle\pi_{n,n+1}\rangle = (-1)^n \pi$. The resulting single-particle dispersion relation is very similar to that obtained for the continuum model in Eq.~\eqref{cont_disp}:
\begin{align}
    \varepsilon_{\text{MF}}(k) = \pm \sqrt{4\sin^2\del{k/2} + g_x^4 \sigma^2 \cos^2\del{k/2} + g_y^4 \pi^2  }\,,
\end{align}
and leads to the following effective potential for $\sigma$ and $\pi$ :
\begin{align}
	V_{\text{eff}}^L(\sigma,\pi) = \frac{\Braket{H}}{N_s N} = \frac{g_x^2}{2}\sigma^2 +\frac{g^2_y}{2} \pi^2 - \int_{-\pi}^{\pi} \frac{\mathrm{d}k}{2\pi} \sqrt{4\sin^2(k/2) + g_x^4\sigma^2 \cos^2(k/2) + g_y^4 \pi^2 }\,,
	\label{effpot MF}
\end{align}
where $N_s$ is the number of lattice sites. In contrast to the continuum effective potential in Eq.~\eqref{effpot_cont}, the lattice effective potential is never invariant under continuous chiral rotations, i.e.\ rotations in the $(\sigma,\pi)$ plane. As we will now argue, this has some non-trivial implications. Most notably, we will find that the absence of continuous chiral symmetry leads to a different mean-field phase diagram on the lattice as in the continuum.
\begin{figure}
	\centering
	\subfigure{\includegraphics[width=0.49\textwidth]{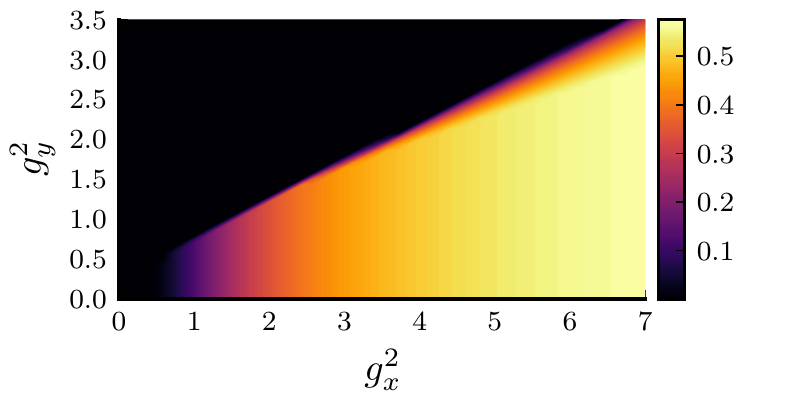}}
	\subfigure{\includegraphics[width=0.49\textwidth]{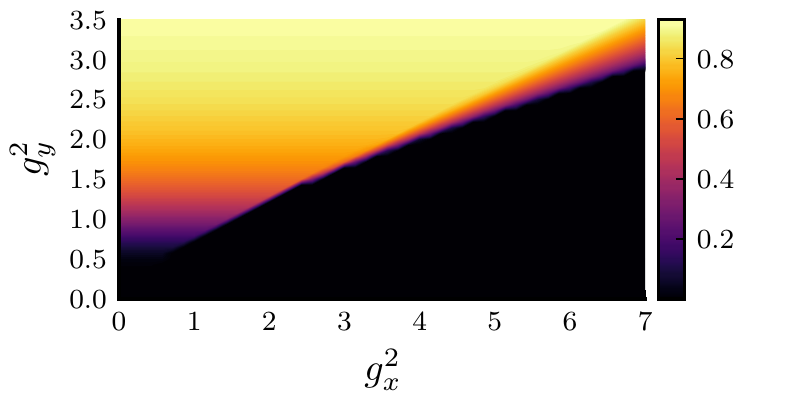}}
	\caption{The value of the order parameters $\sigma$ (left) and $\pi$ (right) that minimize the large-$N$ effective potential $V^L_{\text{eff}}(\sigma,\pi)$ as a function of the two couplings $g_x^2$ and $g_y^2$. 
	}
	\label{MF_order}
\end{figure}
\begin{figure}
    \centering
    \includegraphics[width = 0.49\linewidth]{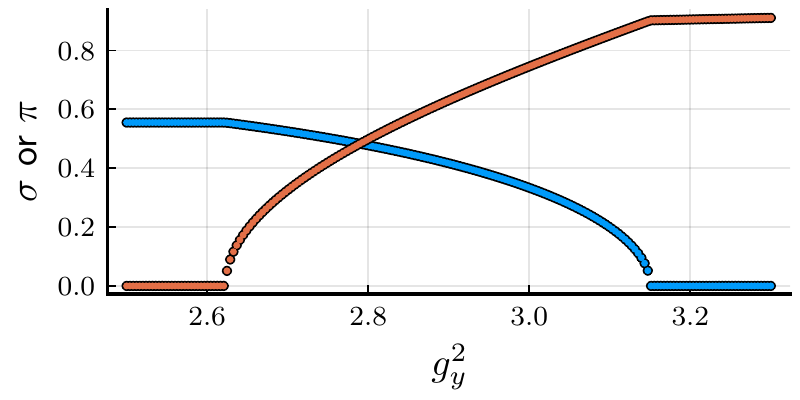}
    \includegraphics[width = 0.49\linewidth]{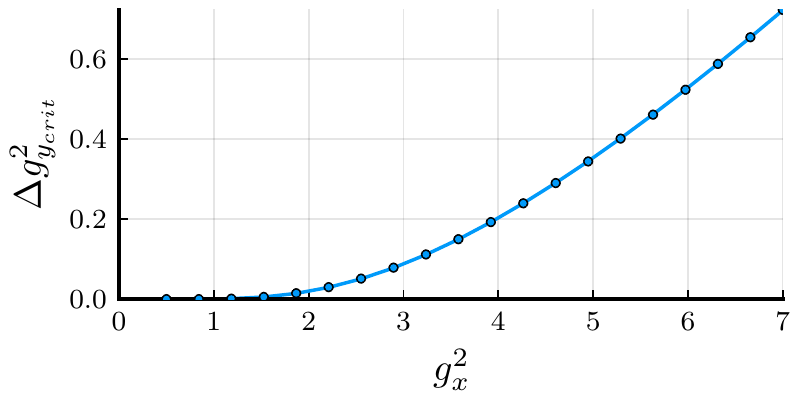}
    \caption{The left panel depicts the behavior of the order parameters $\sigma$ and $\pi$ throughout the phase transitions for fixed $g_x^2 = 6$. Here we can clearly see that there is a coexistence phase where both order parameters are nonzero. The right panel plots the width of this coexistence region as a function of $g_x^2$. The fitted curve is of the form $\Delta g_y^2 \propto g_x^2\exp(-2\pi/g_x^2)$.}
    \label{mf_cut}
\end{figure}

For $g_x^2 \gg g_y^2$, $V_{\text{eff}}^L$ is shown in the left panel of Fig.~\ref{effective_potential}, where we find two mimima along the $\pi=0$ axis. Increasing $g_y^2$ eventually brings us into a coexistence region where both $\sigma$ and $\pi$ are non-zero, corresponding to four distinct minima in the effective potential as shown in the central panel of Fig.~\ref{effective_potential}. Further increasing $g_y^2$ gradually moves the four minima towards the $\sigma=0$ axis and eventually causes them to merge in pairs on said axis. The resulting mean-field phase diagram is shown in Fig.~\ref{MF_order}. In Fig.~\ref{mf_cut}, we plot both $\sigma$ and $\pi$ along a cut of constant $g_x^2 = 6$. This plot clearly shows the coexistence region where both $\sigma$ and $\pi$ are non-zero. Fig.~\ref{mf_cut} also reveals that both $\sigma$ and $\pi$ change continuously as a function of $g_y^2$, which implies that the coexistence region is bounded by two mean-field Ising transitions. 

In the second panel of Fig. \ref{mf_cut} we plot the width of the coexistence region along $g_y^2$ as a function of $g_x^2$. Interestingly, we find that this width becomes extremely narrow for small couplings. In particular, the width decays exponentially according to $\Delta g_y^2 \propto g_x^2e^{-2\pi/g_x^2}$. This suggests that fluctuations beyond mean-field theory can have a non-trivial effect on the phase diagram. Generically, the only effect of fluctuations on a continuous mean-field transition consists of slightly shifting the location of the transition and changing the critical exponents. Here, however, because we have two mean-field transitions that are exponentially close to each other in parameter space, it is conceivable that quantum fluctuations can cause them to merge into a single transition. One reason to expect this is that quantum fluctuations generically tend to restore the symmetry and thus increase the extent of the symmetric (i.e.\ disordered) phase in favor of the symmetry broken phase. Applied to our setting, this implies that the two phase boundaries of the coexistence region, which are already exponentially close in mean-field theory, will be pushed even closer together by the quantum fluctuations. In the following section, we will simulate the $N = 2$ GGN model with MPS and show that the coexistence region indeed disappears in favor of a direct continuous transition.

To summarize, we have found that, although a large-$N$ or mean-field analysis can be used to correctly capture the physics of dynamical mass generation in the GN model (both in the continuum and on the lattice), near the line with continuous chiral symmetry one is nevertheless forced to go beyond mean-field theory. In the continuum, quantum fluctuations are necessary to restore the broken continuous chiral symmetry, whereas on the lattice these same fluctuations are required to merge the two mean-field Ising transitions into a single $c = 1$ CFT. It is interesting that even though the quantum fluctuations play a different physical role in the continuum and on the lattice, they ultimately give rise to the same physics.

\section{Matrix product state simulations at $N=2$}
\label{MPS simul}

This section presents the results of our numerical simulations of the $N=2$ GGN model with tensor networks. We use matrix product states (MPS) \cite{Schollwoeck2011} as a variational class of states for approximating the ground state of the lattice Hamiltonian in Eq~\eqref{latticeH} at different values for ($g_x^2,g_y^2$). More specifically, we work with infinite MPS with a two-site unit cell and use the VUMPS algorithm \cite{ZaunerStauber2018} to find a variationally optimal ground-state approximation directly in the thermodynamic limit. We have explicitly encoded the $\mathrm{SU}(2)\otimes\mathrm{U}(1)$ symmetry into the MPS tensors, allowing us to reach much higher accuracy.\footnote{Our implementation of the MPS algorithms can be found in the Julia package ``MPSKit.jl'' \cite{mpskit}, whereas the (non-abelian) symmetric tensor operations are performed using the ``TensorKit.jl''\cite{tensorkit} package.} The only approximation in our simulations comes from the finite MPS bond dimension $D$, which controls the amount of quantum fluctuations that are taken into account. The bond dimension $D$ corresponds to a truncation of the Schmidt spectrum at a certain treshold $\epsilon$ along any cut in the MPS. In our simulations, we set this truncation threshold $\epsilon$ to a fixed value (which indirectly determines $D$), extract an effective length scale associated to this truncation \cite{Rams2018, Vanhecke2019}, and use this scale to extrapolate our results to the infinite-$D$ limit. We estimate the error on the extrapolation as the change in its value when the highest bond dimension ground state is discarded from the extrapolation procedure. For more details concerning the numerical procedures, we refer to our previous paper, where we applied the same MPS techniques to the conventional GN model \cite{Roose2020}.

\begin{figure}[h!]
	\centering
	\subfigure{\includegraphics[width=0.49\columnwidth]{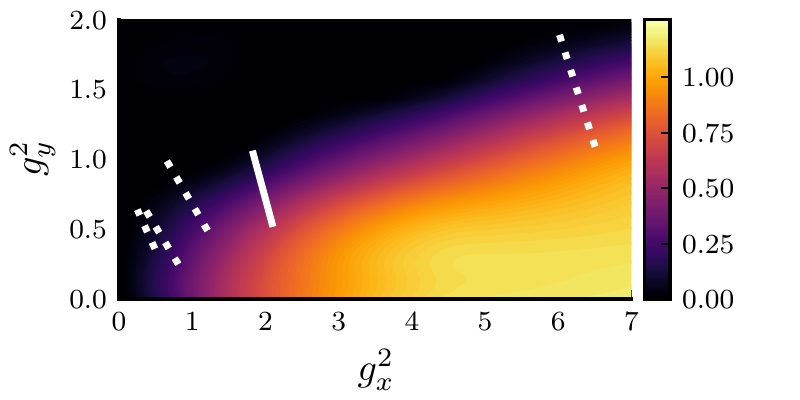}}
	\subfigure{\includegraphics[width=0.49\columnwidth]{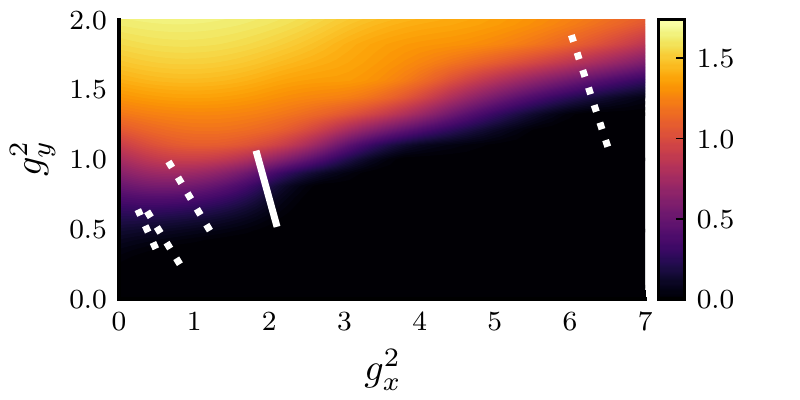}}
	\caption{The phase diagram as computed with infinite MPS with truncation error $\epsilon \approx 10^{-4}$. We show the expectation value of the $\sigma$ (left) and $\pi$ (right) order parameters as a function of $g_x^2$ and $g_y^2$. The full white line indicates the cut that we will analyse in detail below, the dotted lines depict lines we used to study the scaling of $K$ towards the continuum limit.  }
	\label{fig:order}
\end{figure}

To get a first rough idea of the location of the phase transition for $N=2$ we have scanned the parameter space using MPS with truncation error of the order $\epsilon\approx 10^{-4}$. The corresponding bond dimensions range from $D\sim 10$ for points far from criticality to $D\sim 120$ for points close to criticality. The resulting approximate phase diagram is shown in Fig.~\ref{fig:order}. We clearly find two large different regions characterized by either $\sigma = \langle \bar{\psi}\psi\rangle \neq 0$ and $\pi = \langle \bar{\psi}i\gamma_5\psi\rangle = 0$ or $\sigma = 0$ and $\pi \neq 0$. For small these low values of the MPS bond dimension, we also find a small coexistence region, where both expectation values are nonzero, in line with the lattice mean-field results from the previous section.

\begin{figure}[h!]
	\centering
	\subfigure{\includegraphics[width=0.49\columnwidth]{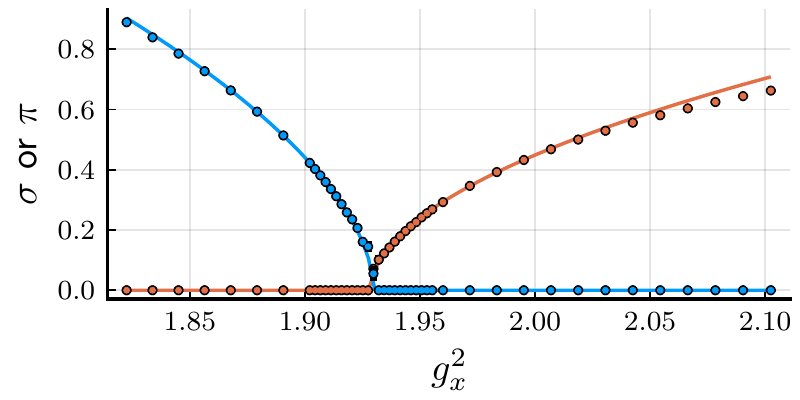}}
	\subfigure{\includegraphics[width=0.49\columnwidth]{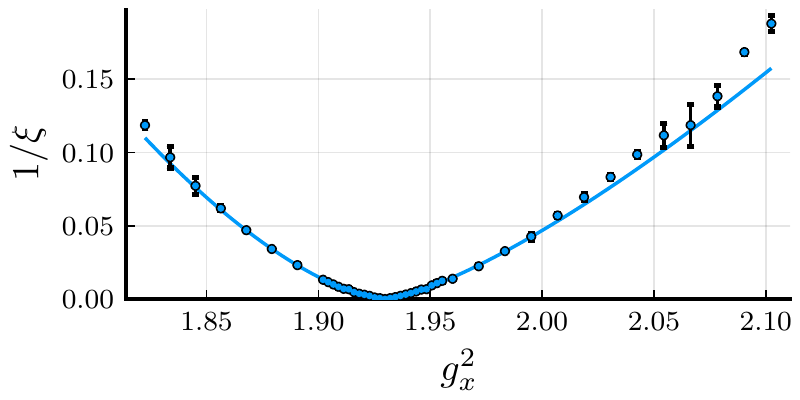}}
	\caption{(left) The extrapolated order parameters $\sigma$ and $\pi$ (plotted in orange and blue respectively) as $g^2_x$ and $g^2_y$ are tuned to take us trough the transition at $g_x^2 \approx 2$. (right) The extrapolated inverse correlation length for the same couplings. The highest bond dimensions used for these simulations are of the order $1100$. The continuous lines represent four independent power-law fits to the numerical data (for $\sigma$, $\pi$, and for both sides of $\xi^{-1}$ separately).}
	\label{RL_cut}
\end{figure}

Let us now focus on the phase transition region and check whether the two mean-field Ising transitions and the coexistence region in between survive as we take more quantum fluctuations into account, by increasing the MPS bond dimension, or whether these transitions actually merge into a single continuous transition. In the left/right panel of Fig.~\ref{RL_cut} we respectively show the two order parameters/inverse correlation length ($1/\xi$) as $g_x^2$ is tuned along the (non-dotted) white line from Fig.~\ref{fig:order}. To each of these quantities, we fit a power law and extract a value for the critical point, resulting in values 1.929 (for $\sigma$), 1.930 (for $\pi$) and 1.929 (for $1/\xi$) that agree reasonably well, thus indicating a direct transition.

To further confirm the scenario of a direct transition, we now try to verify that, close to the critical line, we recover a compact boson theory in the infrared so that the transition has central charge $c=1$. For compact bosons, the scaling dimensions of operators $e^{i n \theta}$ are well known (see e.g.\ Ref.~\cite{Giamarchi:743140}), and are given by $\Delta_n = \frac{n^2 K}{4}$. The scaling dimensions of the operators relevant to our discussion, i.e.\ the operator driving the phase transition ($\cos 2\theta$) and the order parameters ($\cos \theta$ and $\sin\theta$), are respectively $\Delta_{\text{pert}} = K$ and $\Delta_{\text{order}} = K/4$. From this we find that the critical exponents for the correlation length $\nu$ and order parameter $\beta$ are
\begin{align}\label{critical_exponents}
    \nu = \frac{1}{2-K}\;\;\;\;,  \hspace{1 cm}\beta = \frac{K}{8-4K}\, .
\end{align}
Using these relations, we obtain four different estimates for $K$ corresponding to the critical exponents $\nu$ for both order parameters and two critical exponents for the correlation length, i.e.\ one for either side of the phase transition. The four values for $K$ we obtain in this way are respectively $K = 1.351$, $1.381$, $1.351$ and $1.267$, and agree reasonably well with each other. 

\begin{figure}[h!]
	\centering
	\subfigure{\includegraphics[width=0.44\columnwidth,trim={0 0 0.0cm 0},clip]{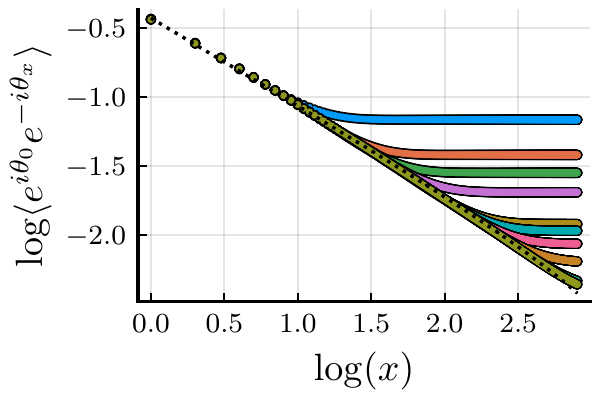}}
	\subfigure{\includegraphics[width=0.44\columnwidth,trim={0 0 0.0cm 0},clip]{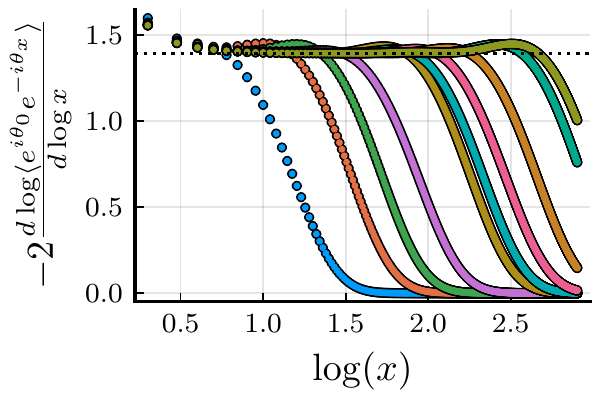}}
	\subfigure{\includegraphics[width = 0.01\columnwidth, height = 4.2cm, trim={0 -0.4cm 0.89cm 0},clip]{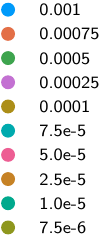}}
	\subfigure{\includegraphics[height = 4.2cm, trim={0.41cm -0.4cm 0 0},clip]{./Figures/legende.pdf}}
	\caption{(left) The two-point correlator of the $e^{i\theta}$ field at transition for the data point closest to the transition. The exponent for the fitted decay is $K\approx1.257$ consistent with all previous estimates. (right) The logarithmic derivative of the two point function, using a finite difference estimator of the data in the left panel. For the larger bond dimensions we see a plateau spanning over roughly 80 sites; the value of this plateau coincides with $K$. In both figures the different colors correspond to decreasing values of the MPS truncation threshold, which are shown on the right.}
	\label{RL_cut_tp}
\end{figure}

Extracting $K$ via the scaling of the order parameter or correlation length is numerically very costly due to the fact that we need many data points close to the transition to accurately fit the critical exponents. Alternatively, we can also obtain $K$ directly from the two-point function of the operators $e^{i\theta} = \frac{\sigma + i \pi}{\sqrt{\sigma^2 + \pi^2}}$ at the critical line, which at large distances should fall off as
\begin{equation}
\langle e^{i\theta(x)} e^{-i\theta(x')}\rangle \sim \frac{1}{|x-x'|^{K/2}} \,.
\end{equation}
In the left panel of Fig.~\ref{RL_cut_tp} the two-point function of the data point closest to the extrapolated critical coupling is shown. The different colors correspond to decreasing values of the MPS truncation threshold. We have fitted a power law to the data with the highest bond dimension, and find a value $K\approx1.257$, again consistent with the previous methods.

In order to make the algebraic decay more clear from the MPS data, in the right panel of Fig.~\ref{RL_cut_tp} we show the logarithmic derivative of the two-point function, i.e. 
\begin{equation}
\eta(x) := -2 \frac{\mathrm{d}\left(\log|\langle e^{i\theta(0)}e^{-i\theta(x)}\rangle|\right)}{\mathrm{d}\log x}\,,
\end{equation}
where again the different colored points correspond to increasing bond dimensions. For the smallest bond dimensions $\eta(x)$ is monotonically decreasing, corresponding to faster than algebraic decay. However, for the larger bond dimensions we can clearly identify a range where $\eta(x)$ is constant, corresponding to a range of algebraic decay; the value for $K$ can now simply be be read off as the value of $\eta(x)$ at the plateau. We estimate the error for $K$ by considering the standard deviation $\sigma_K$ away from the plateau value $\eta_p$, i.e. $\sigma_K^2 =  \langle (\eta(x) - \eta_p)^2 \rangle$, calculated using the data points near the centre of the plateau. Using the plateau in $\eta(x)$ obtained at the largest bond dimensions, we find $K = 1.3944 \pm 0.0006$. Note that the value for $K$ extracted from $\eta(x)$ ($K = 1.394)$ is slightly different from the value we previously obtained via the direct fit in the leftmost plot of Fig.~\ref{RL_cut_tp} ($K  = 1.257$). The value obtained from $\eta(x)$ is less prone to fitting errors, and it also agrees better with the previous estimates for $K$ based on the scaling behaviour of the order parameters and the correlation length.

\begin{figure}[h!]
	\centering
    \includegraphics[width=0.49\columnwidth]{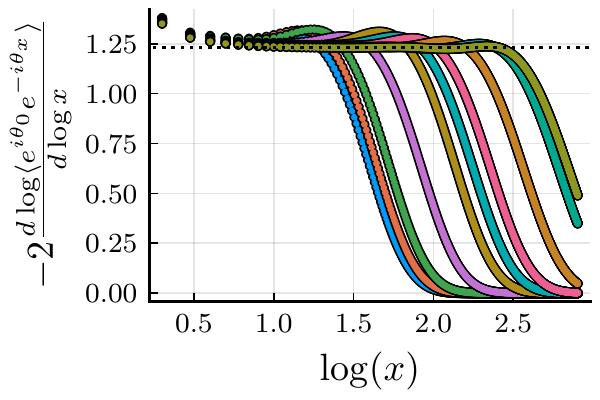}
	\includegraphics[width=0.49\columnwidth]{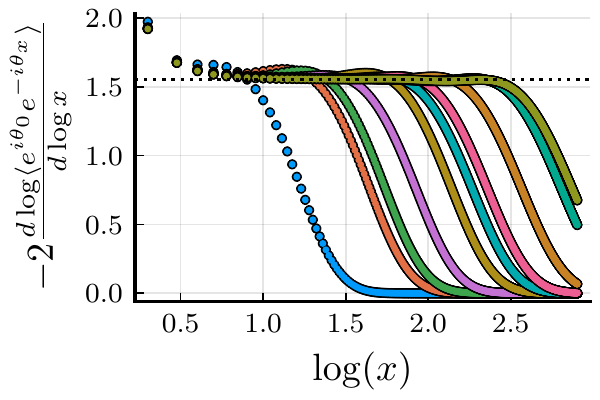}
    \includegraphics[width=0.49\columnwidth]{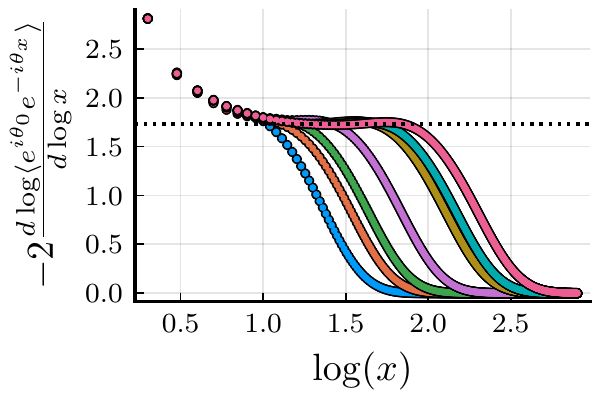}
    \includegraphics[width=0.49\columnwidth]{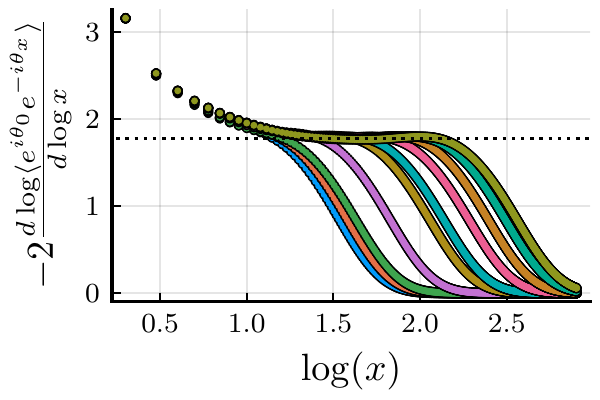}
	\caption{The logarithmic derivative of the two-point function for various points close to the phase transition. The dotted lines indicate the value of the plateau and the corresponding extrapolated value of $K$. The cuts correspond to $g_x^2 \approx 6$ (top-left), $1$ (top-right), $0.6$ (bottom-left) and $0.5$ (bottom-right). The bond dimensions used in these simulations range up to 2500 for the $g_x^2\approx0.5$ point. The color coding is the same as in Fig.~\ref{RL_cut_tp}}
	\label{plateaus}
\end{figure}
To find the critical point and $K$ along the remaining cuts we estimate the position of the critical point using the order parameters, and estimate $K$ by identifying the plateaus at the obtained transition point. These plateaus are shown in Fig.~\ref{plateaus}.

\begin{figure}[h!]
	\centering
    \includegraphics[width=9cm]{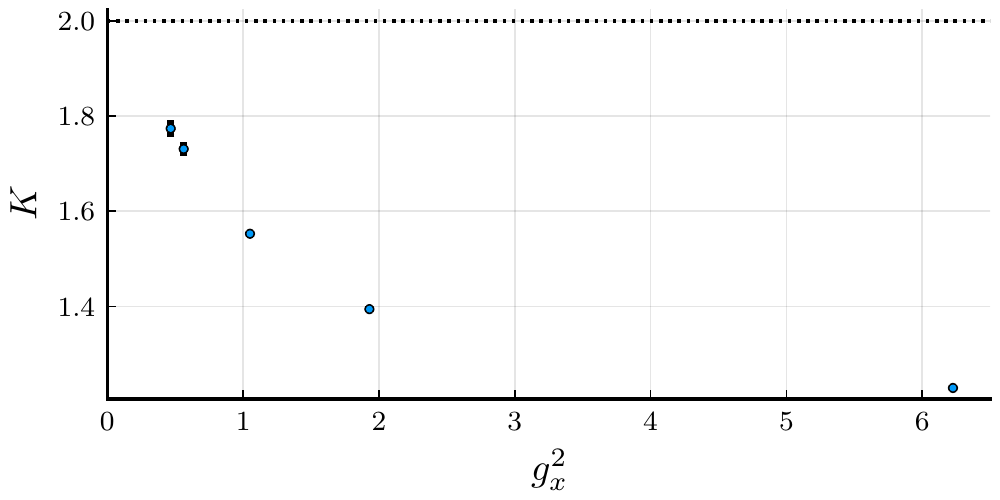}
	\caption{$K$ as a function of the coupling $g_x^2$ for small; we also show the error measure $\sigma_K$ as defined in the text. The dotted line indicates the expected continuum behaviour.}
	\label{K_x}
\end{figure}
Finally, in Fig.~\ref{K_x} we show the values of $K$ obtained from the plateaus in figure~\ref{plateaus} as a function of the coupling $g_x^2$. For sufficiently small values of this quantity we find a clear trend towards the desired continuum value $K = 2$.

\section{The chiral GN model as a Landau-forbidden phase transition}

In the previous section, we have shown that we can recover the behavior of the continuum GGN model at $N= 2$ on the lattice, despite the absence of an exact microscopic continuous chiral symmetry. In this section, we will interpret this result from a condensed matter point of view, and discuss the connection to Landau-forbidden phase transitions. 

We will again focus on the region of parameter space close to where we recover the chiral GN model. The IR physics is then described by the compact boson action in Eq.~\eqref{IR_action}. The discrete symmetries of the GGN model that act non-trivially on the compact boson $\theta$, as discussed in Sec.~\ref{reviewGN}, are:

\begin{eqnarray}
    \mathbb{Z}_2^{\mathcal{D}} & : & \theta \rightarrow \theta + \pi \\
    \mathbb{Z}_2^{\mathcal{C}} & : & \theta \rightarrow -\theta \\
    \mathbb{Z}_2^{\mathcal{R}} & : & \theta \rightarrow -\theta.
\end{eqnarray}
The value of $K$ in Eq.~\eqref{IR_action} is such that the $\cos 2\theta$ operator is relevant, which means that the IR fixed point is indeed a compact boson only when $\delta = 0$. For $\delta>0$, the cosine term will pin $\theta$ to either $\pi/2$ or $-\pi/2$, such that $\mathbb{Z}_2^{\mathcal{D}}$, $\mathbb{Z}_2^{\mathcal{C}}$ and $\mathbb{Z}_2^{\mathcal{R}}$ are all spontaneously broken. However, for $\delta >0$ the ground states are still symmetric under the products $\mathcal{DR}$ and $\mathcal{DC}$ (which act non-trivially on $\theta$). For $\delta<0$, the cosine term will pin $\theta$ to either $0$ or $\pi$, in which case $\mathbb{Z}_2^{\mathcal{D}}$ is spontaneously broken, but $\mathbb{Z}_2^{\mathcal{C}}$ and $\mathbb{Z}_2^{\mathcal{R}}$ are preserved.

\begin{figure}[h!]
	\centering
	\includegraphics[width=5cm]{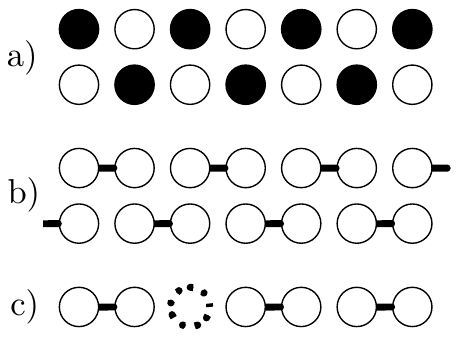}
	\caption{A cartoon picture of zero correlation length representatives of the two symmetry broken phases. a) represents the two charge density waves that occur for large $g_y^2$ ($\pi \neq 0)$. Full/empty dots represent filled/empty sites. b) represents the two different ground states with bond order for large $g_x^2$ ($\sigma \neq 0$). The connected dots represent the dimerized states $(|10\rangle + |01\rangle)/\sqrt{2}$ in the fermion occupation basis. c) represents a kink in the $\sigma$ condensate. Imposing inversion symmetry around the central site automatically breaks all possible inversion symmetries around bonds, which ensures that any such defect nucleates a non-zero value of the $\pi$ order parameter.}
	\label{drawing_kink}
\end{figure}

On the lattice, the reflection operator $\mathcal{R}$ corresponds to a \emph{bond-centered} reflection $\mathcal{R}_B$ (which also includes an on-site action), as discussed in Section~2.3. The $\mathcal{DR}$ symmetry, on the other hand, is realized on the lattice as a \emph{site-centered} reflection $\mathcal{R}_S$, i.e.\ the reflection center now coincides with a lattice site. These two different reflection operators are related by $\mathcal{R}_S = \mathcal{T}\mathcal{R}_B$, where $\mathcal{T}$ is the translation operator, which is consistent with the fact that the latter implements the discrete chiral symmetry on the lattice. The bond-centered reflection symmetry $\mathcal{R}_B$ is broken when $\delta >0$ ($\pi = \langle \bar{\psi}i\gamma_5\psi\rangle \neq 0$), and is preserved when $\delta<0$ ($\sigma = \langle \bar{\psi}\psi\rangle \neq 0$). For the site-centered inversion symmetry, the converse is true, i.e.\ $\mathcal{R}_S$ is preserved when $\delta>0$, and broken when $\delta<0$. That the two different gapped phases indeed respect either the bond- or site-centered reflection symmetry can also be understood intuitively from the fixed-point, i.e.\ zero correlation length, representatives of these two phases. This is shown schematically in Fig.~\ref{drawing_kink}.

The above discussion brings us to the interesting conclusion that the chiral GN model can be interpreted as a continuous phase transition between two gapped phases which break different global symmetries. According to the standard Landau theory of phase transitions, such a continuous transition should be a fine-tuned or multi-critical point, which can only be realized by tuning two independent relevant parameters to zero. Here, we find that this is not the case, and we can go between the two symmetry-broken phases via a single continuous transition, by tuning a single parameter. A natural question is thus what is special about our model that makes a direct transition generic and not fine-tuned. As we will now argue, it is the `t Hooft anomaly which places the GGN model outside the standard Landau theory.

\begin{figure}[h!]
	\centering
	\includegraphics{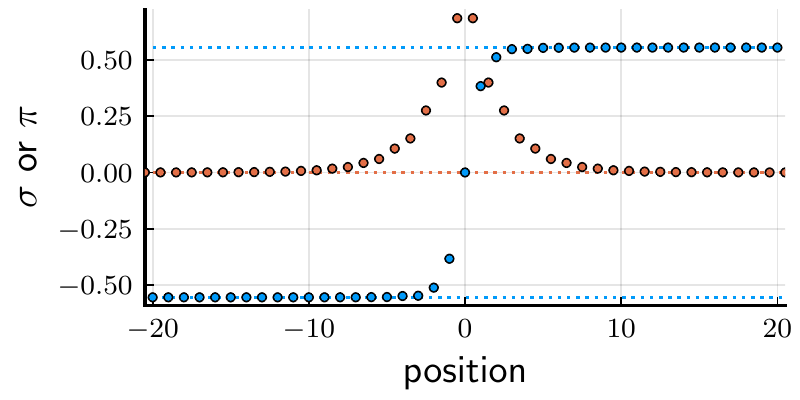}
	\caption{A site-dependent solution of the large-$N$ self-consistency equations with antiperiodic boundary conditions for $\sigma$ (blue), the orange points indicate the $\pi$ order parameter. The couplings are $g_x^2 = 6.0$ and $g_y^2 = 2.3$ ie. in bond ordered phase but close to the transition. Note that the expectation value of $\pi$ becomes nonzero near the domain wall in $\sigma$. \label{sigma_kink} }
\end{figure}

On the lattice, the `t Hooft anomaly between the $\mathbb{Z}_2^{\mathcal{D}}$ and the U(1) charge symmetry is known as the `Lieb-Schultz-Mattis' (LSM) theorem \cite{LSM,OshikawaLSM,HastingsLSM}. It states that at half-filling, the lattice Hamiltonian can only be gapped if either the charge U(1) or the translation symmetry is broken. Note that in our case, with $N=2$ flavors of fermions per site, half filling actually implies that we have one unit of charge per lattice site. We thus also need to invoke the SU(2) flavor symmetry to argue that the average charge is $1/2$ per flavor per site. The LSM theorem then states that a gapped ground state implies that either charge, flavor or translation symmetry are broken. In $1+1$ dimensions, we know from the CHMW theorem that the continuous charge and flavor symmetries cannot be broken spontaneously, so every gapped phase must necessarily break translation symmetry. Let us now assume that we are in a gapped phase where $\mathcal{R}_S$ is broken. A general mechanism to restore the $\mathcal{R}_S$ symmetry is to condense the kink or domain wall excitations. However, because of the relation $\mathcal{R}_S = \mathcal{T}\mathcal{R}_B$, and from the fact that $\mathcal{T}$ must be broken, we conclude that restoring the $\mathcal{R}_S$ symmetry must necessarily imply that we break the $\mathcal{R}_B$ symmetry (assuming that we transition to a gapped phase). This means that condensing the kink excitations must simultaneously \emph{restore} the $\mathcal{R}_S$ symmetry, and \emph{break} the $\mathcal{R}_B$ symmetry. We thus conclude that the `t Hooft anomaly must endow the kink excitations with a special property that their condensation triggers the spontaneous breaking of $\mathcal{R}_B$. In Fig.~\ref{sigma_kink}, we plot the large-$N$ mean-field solution for the ground state of the lattice model with twisted boundary conditions, such that the ground state contains a single kink in the $\sigma$ order parameter. Near the center of the kink, we see that the $\pi$ order parameter becomes non-zero, such that condensing these kinks will induce a uniform non-zero value for the $\pi$ order parameter, signaling a spontaneous breaking of $\mathcal{R}_B$. 

Finally, let us comment on the role of the $\mathbb{Z}_2^{\mathcal{C}}$ symmetry. It turns out that there is a further `t Hooft anomaly, i.e.\ a LSM obstruction on the lattice, between the charge and flavor symmetries, the $\mathcal{C}$ symmetry, and the $\mathcal{R}_S$ symmetry. As we show in the appendix, a ground state of a local and gapped Hamiltonian with an average charge of $1/2$ per flavor and per site cannot be invariant under site-centered reflection symmetry. Because the charge and flavor symmetries cannot be broken due to the CHMW theorem, it thus follows that the $\mathbb{Z}_2^{\mathcal{C}}$ symmetry must be broken in the gapped phase which preserves $\mathcal{R}_S$, i.e.\ when $\delta >0$ such that $\pi = \langle \bar{\psi}i\gamma_5\psi\rangle \neq 0$. We thus again arrive at the conclusion that kink condensation in the $\mathcal{C}$-broken (and also $\mathcal{R}_B$-broken) phase, which restores the charge conjugation symmetry, must necessarily induce $\mathcal{R}_S$ breaking if we are to transition to a gapped phase.

The above discussion of course does not imply that in the presence of the `t Hooft anomalies, there is necessarily a direct continuous transition between a gapped phase with broken $\mathcal{R}_S$ symmetry and a gapped phase with broken $\mathcal{R}_B$ symmetry. It is always possible to have 1) an intermediate region of coexistence where both symmetries are broken, 2) an intermediate gapless region where both symmetries are restored, or 3) a first order transition between the two phases. The `t Hooft anomalies only provide us with a mechanism to explain why a direct continuous transition, if it occurs, is not fine-tuned. This is similar to how the Lieb-Schultz-Mattis theorem is used to motivate the `deconfined quantum critical points' \cite{LevinSenthil,Wang_Chong_Nahum}, which are a special type of Landau-forbidden continuous phase transitions in $2+1$-dimensional lattice spin models \cite{DQCP}.

\section{Conclusion}

In this work we have studied the GGN model on the lattice. In discretizing the GGN model, particular attention was paid to maintaining the maximal amount of global symmetries of the continuum theory. The lattice model used here only fails to preserve the O($2N$) symmetry along the line with $g_x^2 = 0$, and the continuous chiral symmetry, which is present when $g_x^2 = g_y^2$. The latter symmetry has a 't Hooft anomaly, and plays a crucial role in determining the phase diagram of the continuum model. In large-$N$ or mean-field theory, we found that the broken continuous chiral symmetry results in a coexistence region where both $\langle \bar{\psi}\psi\rangle$ and $\langle \bar{\psi}i\gamma_5\psi\rangle$ are non-zero. Interestingly, the width of this coexistence region decreases exponentially with the couplings. To go beyond mean-field theory, we have simulated the $N=2$ lattice GGN model with MPS. We found that the effect of quantum fluctuations beyond mean-field theory is to remove the coexistence region completely, and replace it with a single continuous transition between the phase with $\langle \bar{\psi}\psi\rangle \neq 0$ and $\langle \bar{\psi}i\gamma_5\psi\rangle =0$, and the phase with $\langle \bar{\psi}\psi\rangle = 0$ and $\langle \bar{\psi}i\gamma_5\psi\rangle \neq 0$. The critical line is described by a single compact boson CFT, which is the IR fixed point of the chiral GN model. Although on the lattice the critical line is no longer a straight line under $45^\circ$ in the $g_x^2-g_y^2$ plane (as is the line with continuous chiral symmetry in the continuum model), we found from our MPS simulations that the scaling behaviour of the chiral condensates $\langle \bar{\psi}\psi\rangle$ and $\langle \bar{\psi}i\gamma_5\psi\rangle$ away from the critical line is the same as that predicted by the continuum theory. We have also interpreted the lattice phase diagram from a condensed matter perspective, and explained how the chiral GN model can be recognized as a `Landau-forbidden' phase transition (similar to the deconfined quantum critical point in $2+1$ dimensions) which is not fine-tuned because of the presence of \emph{two} different `t Hooft anomalies, or Lieb-Schultz-Mattis theorems on the lattice. 

This type of continuous transition has been discussed before in the condensed matter literature. Indeed, on the lattice the chiral condensates $\langle \bar{\psi}\psi\rangle$ and $\langle\bar{\psi}i\gamma_5\psi\rangle$ are the order parameters for respectively a bond order density wave or a `Valence Bond State' (VBS) and a `Charge Density Wave' (CDW). The VBS breaks the site-centered reflection symmetry, but preserves the bond-centered reflection and charge-conjugation symmetries. For the CDW, the situation is reversed, i.e.\ it preserves the site-centered reflection symmetry, but breaks the bond-centered reflection and charge-conjugation symmetries. Haldane has found a similar transition in a study of the phase diagram of the antiferromagnetic XXZ chain with next-nearest-neighbour interactions \cite{Haldane} (in the spin language, the CDW corresponds to antiferromagnetic or N\'eel order). This model is closely related to the $N = 1$ GGN model, and it was recently discussed in more detail and generalized in Ref. \cite{Mudry}. In Ref. \cite{SandvikBalents}, the authors studied the phase diagram of the one-dimensional half-filled Hubbard model with an additional nearest-neighbour repulsive interaction using quantum Monte Carlo, and again evidence for a direct continuous transition between a VBS and CDW was found. Because the authors of \cite{SandvikBalents} considered spinful fermions, their Hubbard model is closely related to the $N = 2$ GGN model, although the two interaction terms used in Ref. \cite{SandvikBalents} are different from the ones we obtained here by directly discretizing the continuum GGN model. Another place where a direct continuous transition between VBS and CDW phases has been found (again using quantum Monte Carlo) is the Su-Schrieffer-Heeger model \cite{Weber}. This is a model of fermions hopping on a chain coupled to phonons, and its connections to the GN model were discussed early on in Ref. \cite{FradkinHirsch}. The work presented here makes the connection between the continuous VBS-CDW transition and the GGN model more explicit, as we start by directly discretizing the continuum action of the GGN model. In contrast to the above mentioned previous studies of the VBS-CDW transition, we have also emphasized the importance of two different Lieb-Schultz-Mattis theorems for obtaining a direct continuous transition. Recently, the authors of Ref.~\cite{JiangMotrunich1} have constructed a spin Hamiltonian which was shown \cite{JiangMotrunich2} to exhibit a direct continuous transition between a VBS phase and an Ising ferromagnet phase. The different Lieb-Schultz-Mattis theorems present in this spin model and their importance for the Landau-forbidden phase transition were also discussed in great detail \cite{JiangMotrunich1}. 

In the future, it will be interesting to generalize our numerical results to the GGN model with an odd number of Majorana fermions, in which case the kinks bind an odd number of Majorana zero modes \cite{FendleySaleur} and transform as isospinors under the SO$(\tilde{N})$ symmetry group, where $\tilde{N}$ counts the number of Majorana fermions \cite{Witten}. To simulate these kinks with MPS, one can make use of the results of Ref.~\cite{FermionicMPS}, where it was explained how Majorana zero modes are realized in tensor network states. Another interesting direction is of course to generalize our results to $2+1$ dimensional systems, where discretizing a continuum theory with `t Hooft anomalies might provide a route to construct lattice models with a deconfined quantum critical point. Such a construction is highly desirable, as there is currently no conclusive proof for the existence of a direct phase transition between two different symmetry-broken phases in 2+1 dimensions, despite an impressive numerical effort \cite{Troyer,Wiese,Prokofev,Sandvik1,Sandvik2,Sandvik3,Sandvik4,Nahum1,Nahum2,Damle,Melko,Meng,Lauchli}.

\acknowledgments{We acknowledge valuable discussions with Erez Zohar, Mike Zaletel, and Bram Vanhecke. This work has received funding from the European Research Council (ERC) under the European Unions Horizon 2020 research and innovation programme (grant agreements No 715861 (ERQUAF)), and from Research Foundation Flanders (FWO) via grant GOE1520N and postdoctoral fellowships of LV and NB.}

\appendix

\section{Lieb-Schultz-Mattis obstructions from charge and flavor symmetries, charge conjugation and site-reflection symmetry}

In this appendix we show that there exists a Lieb-Schultz-Mattis (LSM) obstruction (i.e.\ a `t Hooft anomaly on the lattice) in the presence of charge, flavor, $\mathcal{C}$ and $\mathcal{R}_S$ symmetry. In particular, we will show that there cannot exist a quantum state which simultaneously satisfies the following two properties: 1) it is the ground state of a local and gapped Hamiltonian describing a quantum many-body system on a one-dimensional lattice, and 2) it respects all the symmetries mentioned above. To show this, we will rely on the fact that the ground state of every local Hamiltonian with an energy gap can be approximated by an injective\footnote{The injectivity property is a technical condition on MPS tensors which we only mention for completeness in this work -- we will not define it in detail. It suffices to mention that the injectivity condition is physically equivalent to the requirement that the MPS is not a macroscopic superposition or a so-called `cat state'.} finite-bond dimension MPS to arbitrary precision \cite{VerstraeteCirac2006,Hastings2007}. So, it remains to prove that there cannot exist an injective MPS which respects the charge and flavor symmetries, the charge conjugation symmetry and the site-centered reflection symmetry. Our proof will make heavy use of the `fundamental theorem of MPS' \cite{Fundthm}, which allows us to express the global symmetry properties of MPS's in terms of local conditions on the constituent tensors. 

Recall that the gapped phases of interest in this work break translation over a single lattice site (which is the lattice version of the discrete chiral symmetry, so it must be broken in the gapped phases). Because our system has a two-site unit cell, we need two different rank-three tensors $\left[A_1\right]^i_{\alpha\beta}$ and $\left[A_2\right]^i_{\alpha\beta}$ to construct the MPS (we call $i$ the physical index of the MPS tensor, and $\alpha$ and $\beta$ the virtual indices). Concretely, the MPS's we are interested in take the form
\begin{equation}\label{mps}
    |\psi\rangle = \lim_{L \rightarrow \infty} \sum_{\{i_j\}} \langle v|A_1^{i_1}A_2^{i_2}A_1^{i_3}A_2^{i_4}\cdots A_1^{i_{2L-1}}A_2^{i_{2L}} |v\rangle|i_1,i_2,i_3,i_4,\cdots,i_{2L-1}, i_{2L}\rangle \, ,
\end{equation}
where the states $|i_j\rangle$ are a basis for the local Hilbert space on site $j$. For injective MPS, the effect of the choice of boundary vector $|v\rangle$ decays exponentially away from the edge, such that in the limit $L\rightarrow \infty$, the state is independent of $|v\rangle$.

%%%%%%
We now use the fact that we can combine the total charge U(1) and flavor SU(2) symmetry in order to apply separate U(1) transformations on the two flavors (i.e.\ the total charge combined with the diagonal elements from SU(2)). We henceforth refer to this as the charge and flavor U(1) symmetries. The fundamental theorem now implies that an injective MPS of the form in Eq. \eqref{mps} can only be invariant under the charge and flavor U(1) symmetry if the following relations hold \cite{Fundthm,MPSrepresentations}:
\begin{eqnarray}
\sum_{j}\left[U_c(\theta)\right]_{ij}A_1^j & = & e^{iq_1\theta} V(\theta)A^i_1\tilde{V}(\theta)^\dagger \label{Uc1}\\ 
\sum_{j}\left[U_c(\theta)\right]_{ij}A_2^j & = & e^{iq_2\theta} \tilde{V}(\theta)A_2^iV(\theta)^\dagger \, ,\label{Uc2}
\end{eqnarray}
where $U_c(\theta)$ is the local unitary symmetry action corresponding to a U(1) rotation over an angle $\theta$ on flavor $c$, and $V(\theta)$ and $\tilde{V}(\theta)$ are invertible matrices acting on the virtual indices, which without loss of generality can be taken to be unitary matrices \cite{Fundthm}. It is straightforward to see that Eqs. \eqref{Uc1} and \eqref{Uc2} are sufficient for the MPS in Eq. \eqref{mps} to be invariant under the U(1) symmetries. The fact that these local conditions are also necessary is not obvious, but has been proven rigorously in the MPS literature \cite{Fundthm,MPSrepresentations}.

Similarly, the fundamental theorem also implies that the MPS in Eq. \eqref{mps} is invariant under the charge conjugation and site-centered reflection symmetries iff the following relations are true \cite{Fundthm,MPSrepresentations}:
\begin{eqnarray}
\sum_{j}\left[M_C\right]_{ij}A_1^j & = & (-1)^{n_1} CA^i_1\tilde{C}^{\dagger} \\
\sum_{j}\left[M_C\right]_{ij}A_2^j & = & (-1)^{n_2} \tilde{C}A^i_2C^{\dagger} \\
\sum_{j}\left[M_R^1\right]_{ij}\left[A_1^j\right]^T & = & (-1)^{m_1} RA^i_1\tilde{R}^{-1} \\
\sum_{j}\left[M_R^2\right]_{ij}\left[A_2^j\right]^T & = & (-1)^{m_2} \tilde{R}A^i_2R^{-1}\,,
\end{eqnarray}
where $n_i,m_i\in\{0,1\}$, and $M_C$ and $M_R^i$ are the local unitary matrices respectively implementing the charge conjugation symmetry and site-centered reflection symmetry on the physical indices; for our specific model the on-site action depends on the site (even or odd). Furthermore, the site-centered reflection also transposes the MPS matrices as a result of the reordering of the lattice sites. The matrices $C$, $\tilde{C}$, $R$ and $\tilde{R}$ are invertible, and $C$ and $\tilde{C}$ can without loss of generality be taken to be unitary. 

To start our proof, we first note that the U(1) symmetries and the charge conjugation satisfy the following commutation relation:
\begin{equation}
    M_CU_c(\theta) = e^{i\theta}U_c(-\theta)M_C
\end{equation}
Using this relation, we can evaluate $\sum_{j}\left[M_CU(\theta)\right]_{ij} A_1^j$ and $\sum_{j}\left[M_CU(\theta)\right]_{ij} A_2^j$ in two different ways. The first way gives us
\begin{eqnarray}
\sum_{j}\left[M_CU(\theta)\right]_{ij} A_1^j & = & (-1)^{n_1} e^{iq_1\theta} \left(V(\theta)C\right)A_1^i\left( \tilde{V}(\theta)\tilde{C} \right)^\dagger \label{mu1} \\
\sum_{j}\left[M_CU(\theta)\right]_{ij} A_2^j & = & (-1)^{n_2} e^{iq_2\theta} \left(\tilde{V}(\theta)\tilde{C}\right)A_2^i\left( V(\theta)C \right)^\dagger \label{mu2}
\end{eqnarray}
The second way of evaluating this expression leads to
\begin{eqnarray}
\sum_{j}\left[M_CU(\theta)\right]_{ij} A_1^j & = & (-1)^{n_1} e^{i(1-q_1)\theta} \left(CV(-\theta) \right)A_1^i\left(\tilde{C}\tilde{V}(-\theta) \right)^\dagger \label{um1}\\
\sum_{j}\left[M_CU(\theta)\right]_{ij} A_2^j & = & (-1)^{n_2} e^{i(1-q_2)\theta} \left(\tilde{C}\tilde{V}(-\theta) \right)A_2^i\left(CV(-\theta) \right)^\dagger\label{um2}
\end{eqnarray}
For injective MPS, equating \eqref{mu1} and \eqref{mu2} with \eqref{um1} and \eqref{um2}, tells us that the following conditions must hold:
\begin{eqnarray}
V(\theta)C & = & e^{iq\theta}CV(-\theta) \\
\tilde{V}(\theta)\tilde{C} & = & e^{i\tilde{q}\theta}\tilde{C}\tilde{V}(-\theta)  \\
q_1 & = & \frac{1 - q + \tilde{q}}{2} \\
q_2 & = & \frac{1 + q - \tilde{q}}{2}
\end{eqnarray}
At this point, we find it convenient to fix the phase of the matrices $V(\theta)$ and $\tilde{V}(\theta)$ by redefining them as $e^{-iq\theta/2}V(\theta) \rightarrow V(\theta)$ and $e^{-i\tilde{q}\theta/2}\tilde{V}(\theta) \rightarrow \tilde{V}(\theta)$ (note that this also implies $q_1 + q/2 - \tilde{q}/2 \rightarrow q_1$ and $q_2 - q/2 + \tilde{q}/2 \rightarrow q_2$), such that the above equations become
\begin{eqnarray}
V(\theta)C & = & CV(-\theta) \\
\tilde{V}(\theta)\tilde{C} & = & \tilde{C}\tilde{V}(-\theta)  \\
q_1 & =  &q_2 = \frac{1}{2}
\end{eqnarray}
From these relations, we conclude that one of two situations is realized. Either $V(\theta)$ contains integer charges $0$ and $\{Q,-Q\}$ ($Q \in \mathbb{N}^+$) and $\tilde{V}(\theta)$ contains half odd-integer charge pairs $\{q/2,-q/2\}$ ($q \in 2\mathbb{N}+1$), or $V(\theta)$ contains half odd-integer charge pairs and $\tilde{V}(\theta)$ contains integer charges.

For the final step in our proof we use that the reflection and U(1) symmetries commute:
\begin{equation}
    M_RU_c(\theta) = U_c(\theta)M_R\,,
\end{equation}
and evaluate $\sum_j \left[U_c(\theta)M_R\right]_{ij}\left[A_1^j\right]^T$ in two different ways, similarly as before. Equating the two different outcomes now produces the following relations:
\begin{eqnarray}
RV(\theta) & = & e^{iQ_R\theta} \tilde{V}^*(\theta)R\\
\tilde{R}\tilde{V}(\theta) & = & e^{iQ_R\theta} V^*(\theta)\tilde{R}
\end{eqnarray}
These equations imply that the charges of $V(\theta)$ are equal, up to a permutation, to the charges of $\tilde{V}(\theta)$ shifted by $Q_R$. If such a $Q_R$ exists, then from our considerations above it follows that it should be a half odd-integer. However, it is not hard to see that the integer charges $0$ and $\{Q,-Q\}$ cannot be obtained by shifting the half odd-integer charges $\{q/2,-q/2\}$ by some overall half odd-integer (provided that there are a finite number of charges, i.e.\ provided that the MPS bond dimension is finite). So we have arrived at an inconsistency, from which we conclude that there cannot exist an MPS which is invariant under all the symmetries.

\bibliography{library.bib}

\end{document}